\documentclass[12pt,preprint]{aastex}

\shorttitle{X-ray spectral features in Mrk 766}
\shortauthors{K. O. Mason et al.}
\begin{document}
\title{The X-ray spectrum of the Seyfert I galaxy Mrk 766:
Dusty Warm Absorber or Relativistic Emission Lines?}
\author{K. O. Mason$^1$, G. Branduardi-Raymont$^1$, P. M. Ogle$^2$, 
M. J. Page$^1$, E. M. Puchnarewicz$^1$, E. Behar$^3$, F. A. C\'ordova$^2$,
S. Davis$^2$,
L. Maraschi$^4$, I. M. McHardy$^5$, P. T. O'Brien$^6$, W. C. Priedhorsky$^7$ 
\& T.P. Sasseen$^2$} 
\affil{$^1$Mullard Space Science Laboratory, Department of Space and Climate
Physics, \\ University College London, Holmbury St Mary, Dorking, Surrey,
RH5 6NT, UK.\\
$^2$ Department of Physics, University of California, Santa Barbara, CA 93106, USA\\
$^3$ Columbia Astrophysics Laboratory, Columbia University, 550 W 120 St, New York, NY 10027, USA\\
$^4$ Osservatorio di Brera, via Brera 28, 20121 Milano, Italy\\
$^5$ University of Southampton, University Road, Southampton\\
$^6$ Physics and Astronomy Dept., University of Leicester, University Road, Leicester,
UK\\
$^7$ Los Alamos National Laboratory, Los Alamos, NM 87545, USA\\
} \email{kom@mssl.ucl.ac.uk}
\received{2001 June 25 }
\begin{abstract}
Competing models for broad spectral features in the soft X-ray spectrum of
the Seyfert I galaxy Mrk~766 are tested against data from a 130~ks 
XMM-Newton observation. A model including relativistically broadened Ly$\alpha$ emission
lines of \ion{O}{8}, \ion{N}{7} and \ion{C}{6} is a better fit to 0.3-2 keV XMM-RGS
data than a dusty warm absorber. 
Moreover, the measured depth of neutral iron absorption lines in the spectrum 
is inconsistent with
the magnitude of the iron edge required to produce the
continuum break at 17-18\AA\ in the dusty warm absorber model. 
The relativistic emission line model can reproduce the broad-band
(0.1-12 keV) XMM-EPIC data with the addition of a fourth line to represent emission from 
ionized iron at 6.7 keV and an excess due to reflection at energies above the
iron line. The profile of the 6.7 keV iron line is consistent with
that measured for the low energy lines. 
There is evidence in the RGS data at 
the 3$\sigma$ level 
for spectral features 
that vary with source flux. The covering fraction of warm absorber gas is
estimated to be ~12\%. Iron in the warm absorber is found to be overabundant 
with respect to CNO compared to solar values. 

\end{abstract}


\clearpage

\section{Introduction}

The combination of the Reflection Grating Spectrometer (RGS) and the high throughput of 
XMM-Newton constitutes a powerful tool for probing the soft X-ray spectrum of bright 
AGN. The increase in spectral resolution has already caused a revision in thinking 
and has revealed a complex spectrum of narrow absorption lines arising in material 
with a range of ionisation states (e.g. IRAS 13349+2438; Sako et al. 2001). The 
XMM-RGS spectra of the Seyfert galaxies MGC-6-30-15 and Mrk~766  
show evidence of broad spectral features (Branduardi-Raymont et al. 2001; 
hereafter BR01; see also Sako et al. 2002) 
and a similar spectrum of MCG-6-30-15 has been recorded using the
grating spectrometers on the Chandra 
observatory (Lee et al. 2001). The nature of these features is controversial. 
Two competing models are:  (1) absorption edges imprinted on
the underlying spectrum by a dusty, warm absorber (Lee et al. 2001), (2)
emission lines of \ion{O}{8}, \ion{N}{7} and \ion{C}{6} broadened by relativistic 
motion in an 
accretion disk around a massive spinning black hole (BR01). 

In this paper we discuss the first results of a long (130 ksec) XMM-Newton observation 
of Mrk~766. This has better statistics than 
the shorter observation of BR01. 
We examine how well the competing models are able to reproduce the 
RGS data. The way in which the spectral features change with source flux is
investigated by comparing the mean spectrum taken during flaring episodes
with that from intervals between flares. We also discuss the nature of the 
warm absorber.  
The spectrum in the full X-ray energy 
range of XMM-Newton is investigated, combining the RGS data with 
those recorded at the same time
by the EPIC CCD spectral imager. We compare the 
low energy spectral features to the $\sim$ 6.7 keV iron line detected using 
EPIC.   

\section{Observations and Data Reduction}

Mrk~766 was observed with all the XMM-Newton instruments for a continuous period 
of 130 ksec beginning 2001 May 21. This is essentially the maximum observing time
possible during an XMM-Newton orbit.
The RGS was used in standard spectroscopy mode 
(den Herder 2001). The EPIC MOS-1 camera was used in timing mode
while MOS-2 and the PN camera were set to small window mode (Turner et al.
2001; Strueder et al. 2001). The OM instrument (Mason et al. 2001) took a series of ultraviolet images of
Mrk~766 in various filters with a cadence of 1300 sec. 

The data were reduced using the standard XMM-Newton Science Analysis Software. 
In this letter we consider the RGS and 
MOS-2 EPIC data only, the latter because the camera and operating mode 
combination currently has the best established calibration among 
those used. 

\section{Analysis and Results}

We detect X-ray variations in Mrk~766 of up to a factor of two on a 
timescale of a few thousand seconds. The light curve of the source in the 1-12 keV band
as seen by the EPIC MOS-2 camera is shown in Figure~1. Analysis does not reveal
any gross energy dependence associated with the variability (section 3.2), which allows us to
co-add the data from the whole observation to study the X-ray 
spectral shape.

\subsection{RGS data}

The summed RGS data are shown in Figure~2. The spectrum is characterised by a number
of strong broad spectral features, peaking at approximately 19\AA, 25\AA, and 34\AA, 
cut into
by numerous narrow absorption lines. Overall the spectrum is similar in appearance to
that reported by BR01, but the statistical quality is significantly higher. This is
because the present observation was approximately twice as long, and because
the average flux level over the observation 
was 50\% brighter
than it was in the data
of BR01.  A segment of the data is plotted on a larger scale in Figure~3, and compared
with the spectrum derived from the BR01 observation. In this wavelength interval,
the median signal to noise ratio per channel
is 19 in the present data, compared to a value of 9 for the BR01 data. The two spectra
are very similar, apart from excess narrow emission in the BR01 data near
16.4 and 17.5 \AA\ which may be due to higher order iron lines. The narrow absorption lines
are unresolved, and have a width that is consistent with the instrumental resolution.

\begin{figure}
\epsscale{0.7}
\plotone{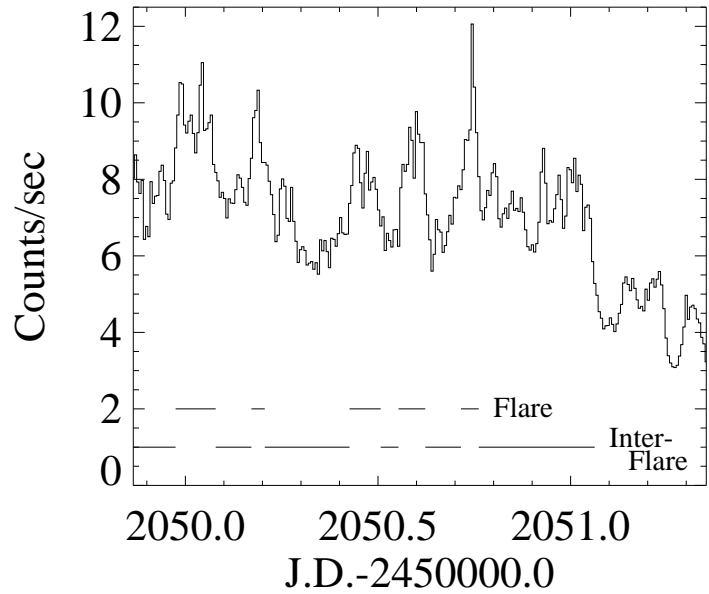}   
\caption{Light curve of Mrk~766 in the 1-12 keV band taken with the EPIC MOS-2 camera
during the current observation. The data are collected in 500s bins. Date intervals
used for later spectral analysis are indicated. These are 
categorised as flaring ('Flare') or between flares ('Inter').}
\end{figure}

\begin{figure}
\epsscale{0.9}
\plotone{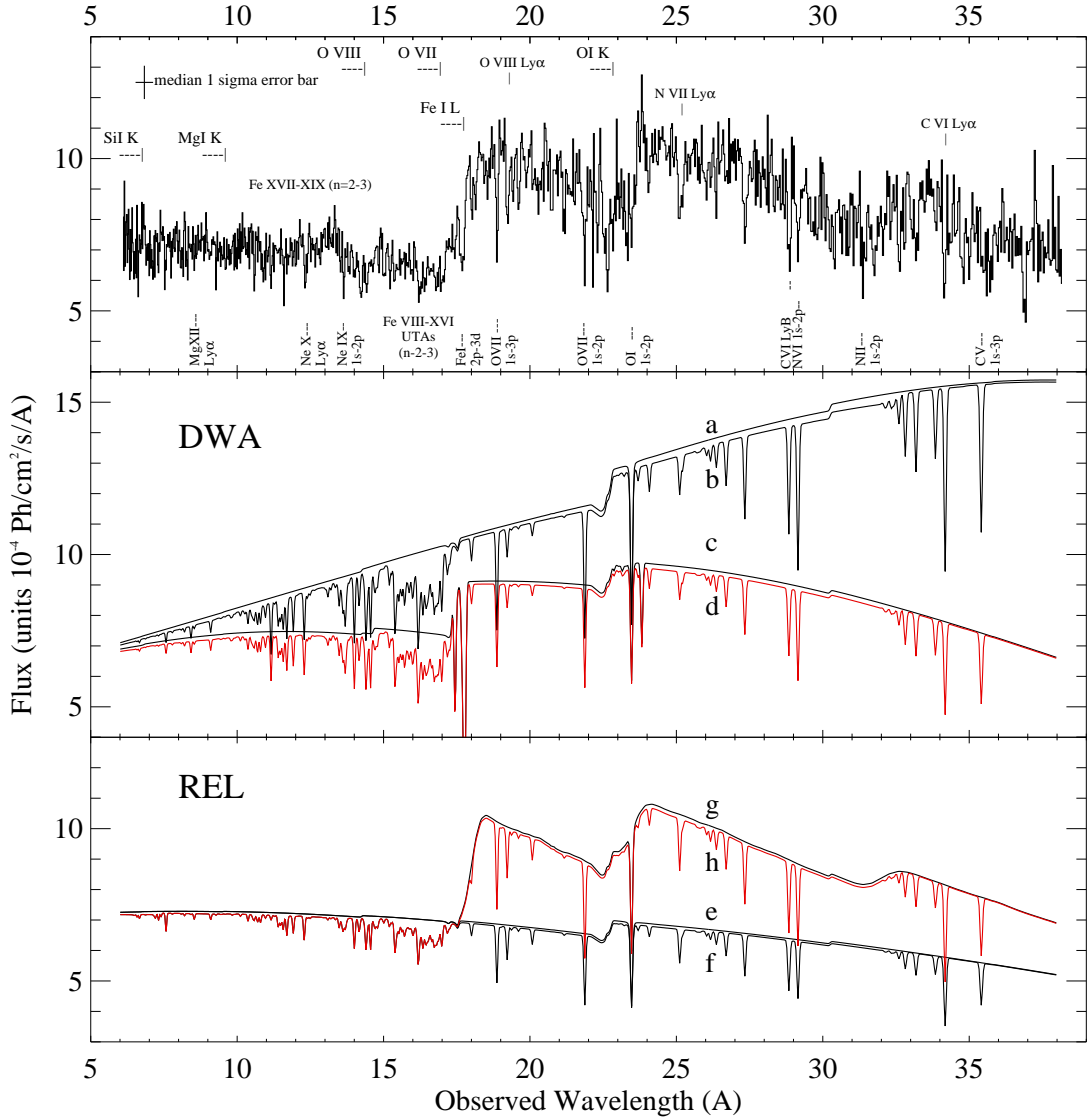}
\caption{{\it Upper panel:} Combined RGS-1 and RGS-2 data (first and second order) 
on Mrk~766 summed over the 130ks 
observation and divided by
the wavelength-dependent effective area of the instrument. The wavelength of 
various features that might be
expected
in the spectrum are marked. In the {\it lower panels} we show the components of the 
DWA and REL models that best fit the data. In each case, the red curve 
(curves {\it (d)} and {\it (h)} respectively
for the DWA and REL case) is the combined best-fit spectrum. Each curve is modified
by Galactic absorption. In the DWA fit, curve {\it (a)} is the combination of two power laws
with photon slopes of 2.1 and 3.7 respectively, curve {\it (b)} includes the power law
continuum and the
warm absorber (dust omitted), while curve {\it (c)} shows the power law continuum
plus dust (warm absorber omitted). In the REL fit, curve {\it (e)} is the single
power law continuum with photon slope 2.15, curve {\it (f)} shows the power law
continuum together with the warm absorber (emission lines omitted) while curve {\it (g)}
shows the power law continuum plus the broad emission lines (warm absorber omitted).
  }
\end{figure}

\begin{figure}
\epsscale{0.9}
\plotone{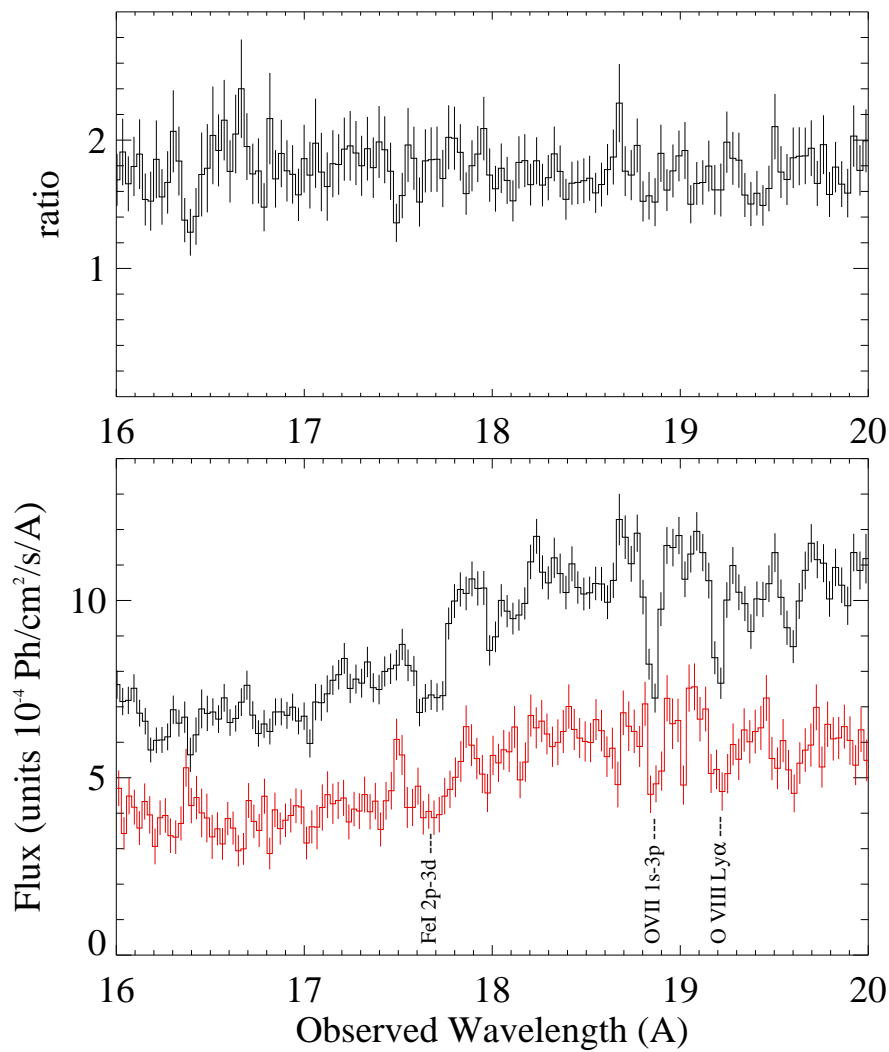}
\caption{A section of the RGS spectrum from the present observation plotted
on an expanded scale (black), compared to the spectrum from the Performance
Verification phase obseration of BR01 (red), reduced with the same version of
the analysis software. The ratio of the two spectra is shown in the upper plot.}
\end{figure}

\begin{figure}
\epsscale{0.7}
\plotone{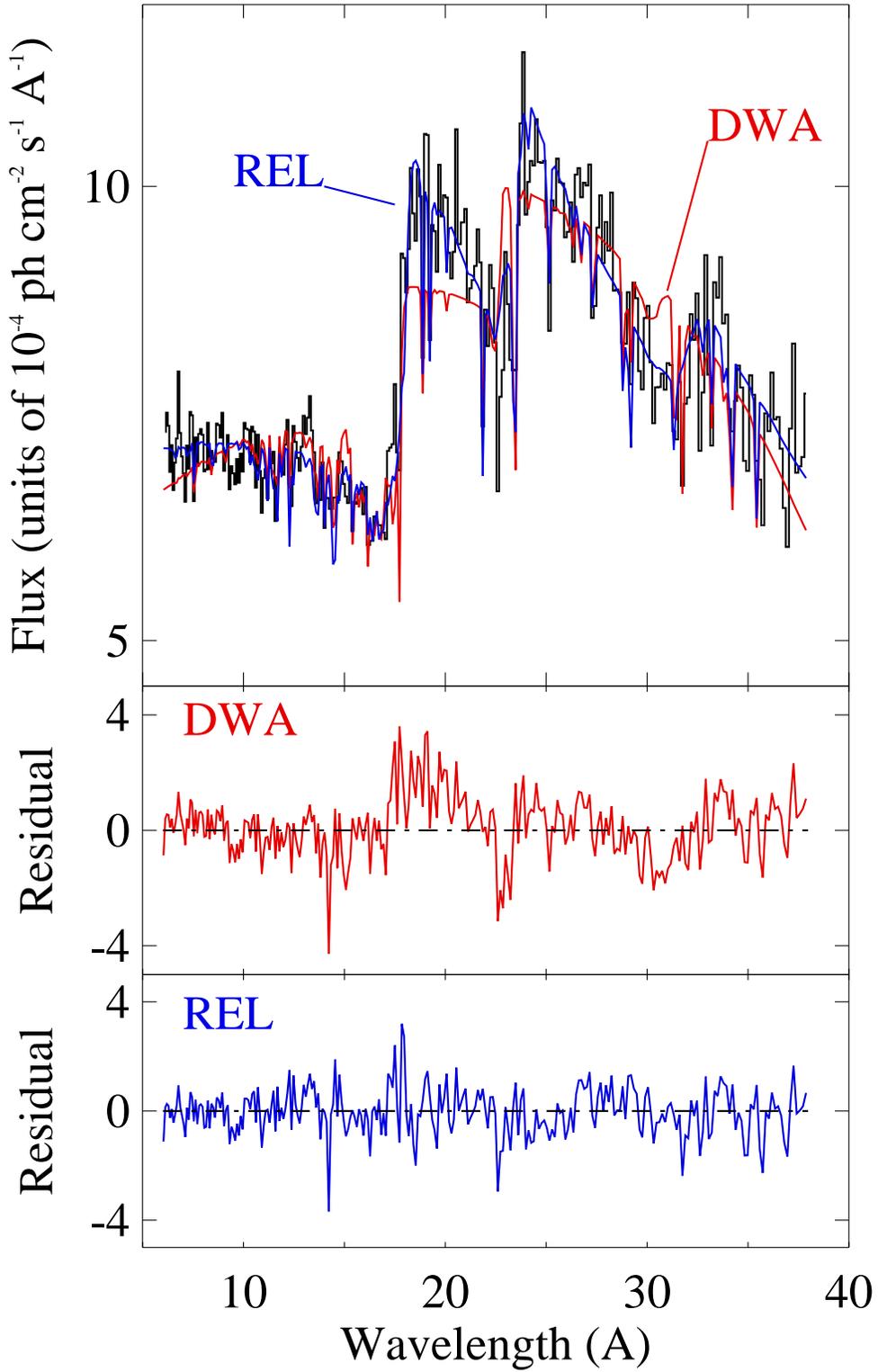}
\caption{Comparison of the DWA (red) and REL (blue) model fits to the RGS data. For clarity of 
display, all data have been binned by a factor of four.}
\end{figure}

\begin{figure}
\epsscale{0.9}
\plotone{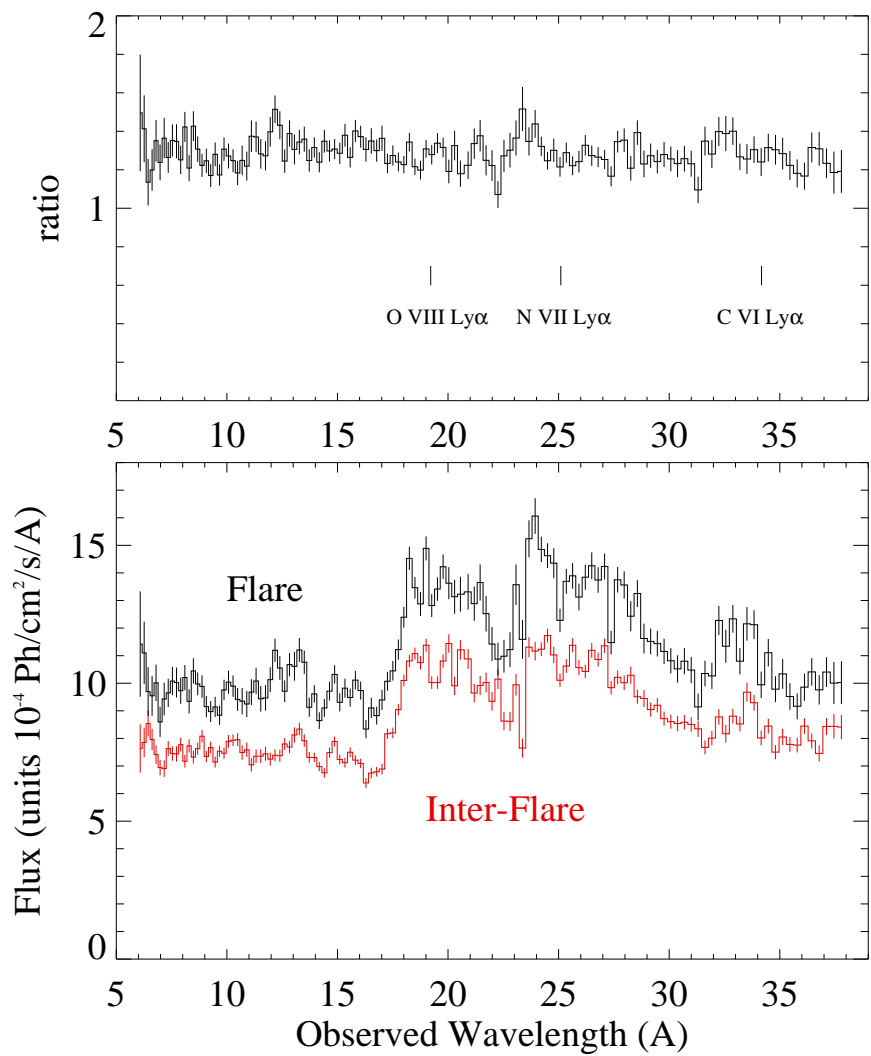}
\caption{The RGS spectrum of Mrk 766 integrated over the peak of flaring
intervals (see Figure 1) compared with the spectrum integrated over 
intervals between flares. The ratio of the two spectra is plotted
in the upper panel. One sigma statistical errors are shown. 
The data have been binned by a factor of eight compared to
Figure~2 to 
improve statistics.}
\end{figure}

\begin{figure}
\epsscale{0.9}
\plotone{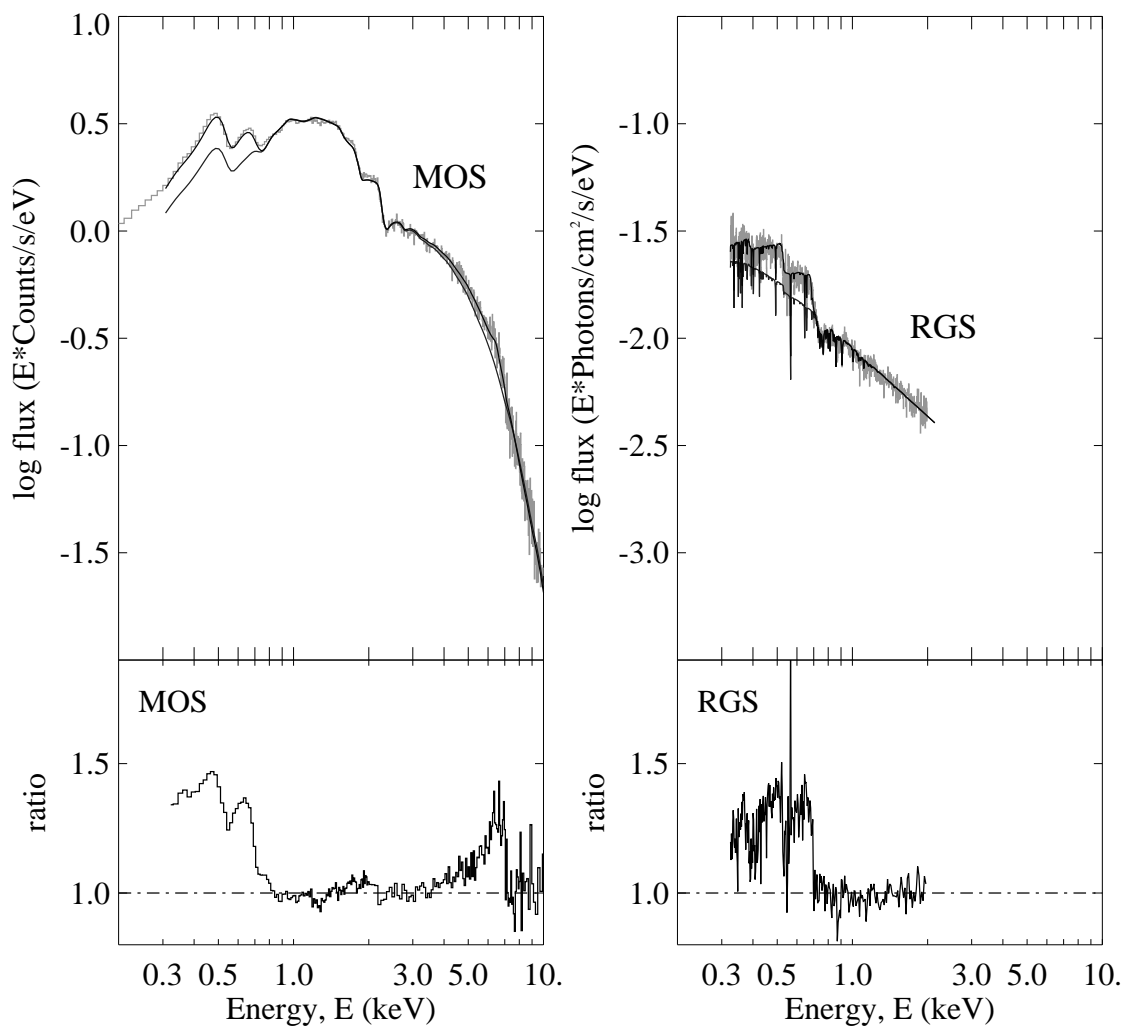}
\caption{{\it Left:} EPIC MOS-2 count spectrum of Mrk~766 showing the best-fit REL model.
{\it Right:}The RGS data from Figure~2 are also plotted, fitted to the
same model. The model with the emission lines suppressed is included for both MOS and RGS to 
illustrate
the contribution of the lines. The ratio of the data to the underlying spectrum
with the lines removed is shown in the lower panels.}
\end{figure}

\begin{figure}
\epsscale{0.6}
\plotone{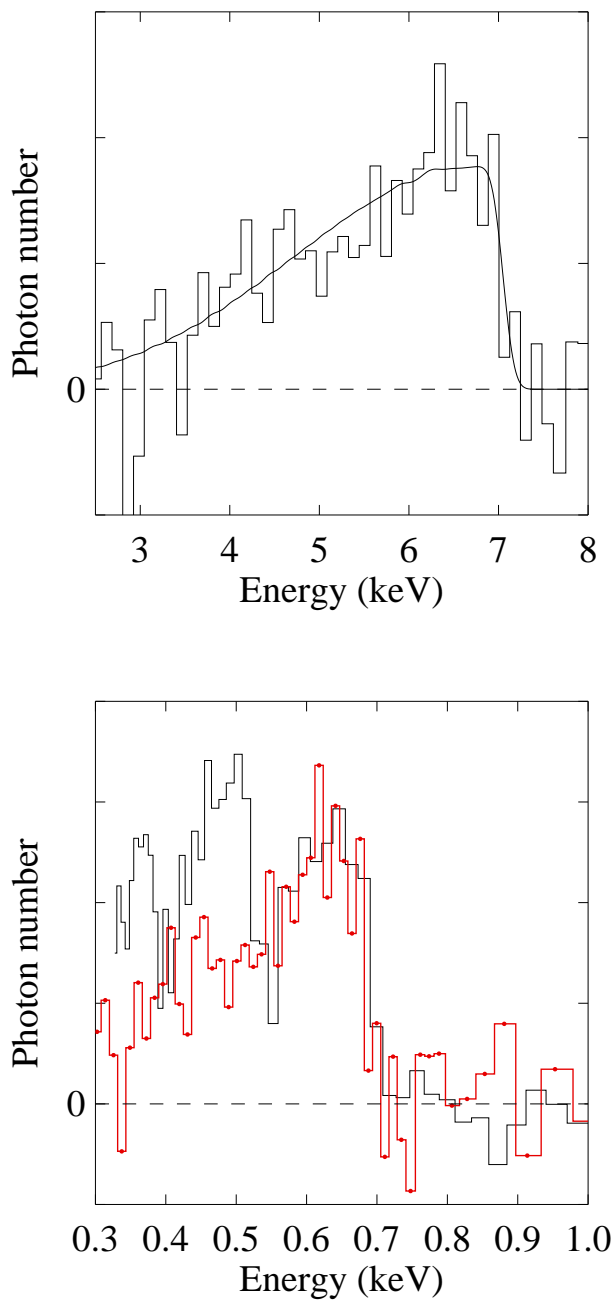}
\caption{The upper panel shows the iron K$_\alpha$  line profile of Mrk~766 derived from the
EPIC MOS-2 data: the histogram shows the residuals from the best fit continuum 
illustrated in Figure~6. The continuum is derived from a model which includes a Laor line
to describe the excess attributed to iron emission, and the best fit Laor profile
is superimposed on the data as the continuous line. In the lower panel
we show the RGS data on Mrk~766 with the best fit continuum subtracted. The red
histogram shows the iron K$_\alpha$ line data from MOS-2 scaled to the energy of
the \ion{O}{8} line. The data have been binned for clarity, and the relative line fluxes
have been normalised to their peak values.
}
\end{figure}
\begin{figure}
\epsscale{0.8}
\plotone{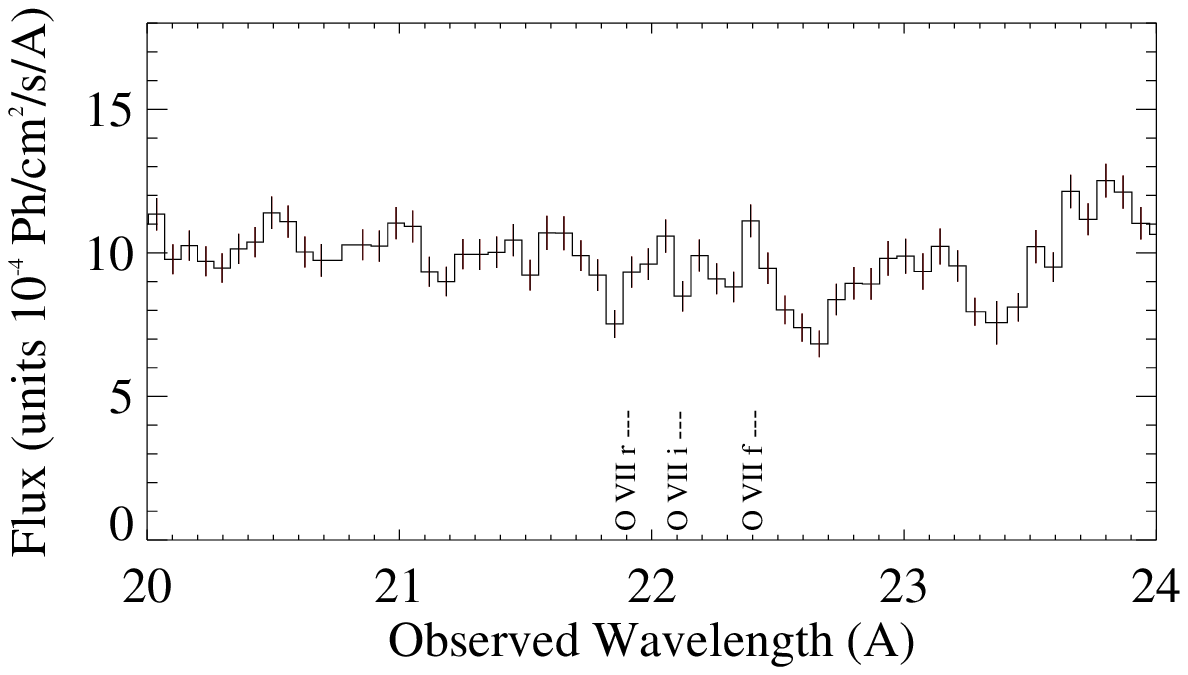}
\caption{An expanded view of the RGS spectrum of Mrk~766 in the vicinity of the
\ion{O}{7} triplet, summed over the whole observation.
}
\end{figure}

We try to fit the spectrum with two different schemes. In the first the broad spectral
features are modelled as edges imprinted on a continuous underlying spectrum by a
combination of neutral atoms in dust and a 
warm absorbing medium, both in the line of sight to the X-ray source. 
In the second the features
are modelled as broad emission lines 
from a relativistic accretion disk, coupled with a foreground warm absorber. Both models
include Galactic absorption with a column density of $1.77 \times 10^{20}$ H atoms cm$^{-2}$. 
We use cross-sections for the Galactic \ion{O}{1} edge computed by McLaughlin \& Kirby (1998)
and include the 
Galactic O I 1s-2p resonance line and
Fe I L 2p-3d resonances. We used data from Behar, Sako \& Kahn (2001) 
for the Fe I L cross sections.

\subsubsection {Dusty Warm Absorber}

\renewcommand{\arraystretch}{.6}

\begin{deluxetable}{lllllllll}
\tablewidth{0pc}
\tablecaption{Ionized Warm Absorber}
\tablehead{
\colhead{Ion}  & \colhead{DWA N$_i$}  & \colhead{REL N$_i$} & 
\colhead{Ion}  & \colhead{DWA N$_i$}  & \colhead{REL N$_i$} &
\colhead{Ion}  & \colhead{DWA N$_i$}  & \colhead{REL N$_i$} \\
}
\startdata
\ion{C}{5} & 3.4 & 2.2     & \ion{Fe}{8} & 0.23 & 0.10 & \ion{Fe}{17} & 0.79 & 0.39\\
\ion{C}{6} & 3.4 & 3.7     & \ion{Fe}{9} & 0.45 & 0.24 & \ion{Fe}{18} & 0.94 & 0.48\\
\ion{N}{6} & 0.58 & 0.42   & \ion{Fe}{10} & 0.72 & 0.43& \ion{Fe}{19} & 0.38 & 0.11\\
\ion{N}{7} & 0.72 & 1.4    & \ion{Fe}{11} & 0.19 & 0.12& \ion{Fe}{20} & 0 & 0\\
\ion{O}{7} & 2.2 & 1.8     & \ion{Fe}{12} & 0.62 & 0.37& \ion{Fe}{21} & 0 & 0\\
\ion{O}{8} & 0.56 & 1.2    & \ion{Fe}{13} & 0.12 & 0.10& \ion{Fe}{22} & 0.57 & 0.30\\
\ion{Ne}{9} & 1.1 & 0.57   & \ion{Fe}{14} & 0.36 & 0.10& \ion{Fe}{23} & 2.0 & 1.5\\
\ion{Ne}{10} & 0.46 & 0.25 & \ion{Fe}{15} & 0.33 & 0.16& \ion{Fe}{24} & 0.86 & 0.46\\
\ion{Fe}{7} & 0 & 0        & \ion{Fe}{16} & 0.29 & 0.16    
\enddata
\tablecomments{units 10$^{16}$ cm$^{-2}$}
\end{deluxetable}

\begin{deluxetable}{ll}
\tablewidth{0pc}
\tablecaption{DWA Neutral Absorber}
\tablehead{
\colhead{Ion}  & \colhead{Column Density (units} \\
\colhead{} & \colhead{10$^{16}$ cm$^{-2}$)}  \\
}
\startdata
\ion{C}{1} & \hfil 97.7$^{+3.3}_{-0.4}$\\
\ion{O}{1} & \hfil 2.4$^{+1.5}_{-1.7}$\\
\ion{Mg}{1} & \hfil 0.0$^{+4.2}_{-0}$\\
\ion{Si}{1} & \hfil 2.3$^{+2.1}_{-0.3}$\\
\ion{Fe}{1} & \hfil 16.5$^{+1.3}_{-0.2}$
\enddata
\tablecomments{errors are 90\% confidence.}

\end{deluxetable}

We have developed a Dusty Warm Absorber (DWA) model that we fit to the Mrk~766 RGS data. 
The model incorporates an ionized (warm)
absorber component, which includes all abundant H-like and He-like ions (Table~1). For
each ion we include K edges and resonance absorption lines from the ground state 
and L-shell absorption from ions Fe VII-XXIV. We 
model the warm absorber empirically, allowing the column density of
each ion to vary so as to obtain a best fit.
The
Doppler b parameter (=$\sqrt{2}\sigma=0.6 FWHM$) of the lines is fixed at an 
assumed 100 km/s. The narrow lines are 
not resolved by the RGS and we can place an upper limit on their directly measured width of b=300 km/s.
We can however constrain the b parameter by comparing the equivalent widths of 
different lines produced by the same ion. The formal best-fit, using the 18-38 \AA\ range 
to avoid the blended iron lines at shorter wavelengths, is 
b=$95 \pm 30$ km s$^{-1}$. There is no evidence for a  
global velocity shift of the lines greater than $\pm 100$ km/s
with respect the systemic velocity (z=0.012929).

We include neutral absorption in the model from the 
K, L and M edges of \ion{Fe}{1} and \ion{O}{1}, and other abundant 
elements that might be present 
including: \ion{C}{1}, \ion{Mg}{1}, \ion{Ne}{1}, and \ion{Si}{1}. Neutral absorption 
edges might arise in dust that is embedded in the warm absorber, as suggested by
Reynolds et al. (1997; see also Komossa \& Bade 1998).

In modelling the H-like and He-like resonance lines up to n=10, we use atomic
data from Verner et al. (1996a). Data for n=11-30 are extrapolated using the 
hydrogenic approximation. The Fe-L shell oscillator strengths and natural line
widths have been computed by Behar et al. (2001). The lines are represented by
Voigt profiles, convolved with the instrument broadening function. The edges are
modelled using cross sections calculated by Verner et al. (1996b).  

Below about 0.7 keV Mrk~766 exhibits an excess above an extrapolation of
the best-fit medium energy power-law (e.g. Page et al. 2001). We thus 
require a soft excess spectral component if we are to fit
the overall spectrum with a DWA model, 
since the DWA itself can only {\it remove} flux. We tried a blackbody 
spectrum to represent this soft excess, but a power law was a much better fit
so we have adopted this in what follows. Galactic absorption with a column
density $N_H = 1.77 \times 10^{20}$ cm$^{-2}$ was included.

The column densities associated with the various warm absorber lines and dust features were
fit along with the continuum parameters so as to minimise $\chi^2$.
The individual components that make up the best fit DWA model are illustrated
in the middle panel of Figure~2. The overall model is compared with the
data directly in Figure~4, and has a $\chi^2$/d.o.f.=2.2 (1094 d.o.f.). 
The counterbalancing
effects of the steep low energy excess and the absorption by neutral atoms
inherent in the model are clear from Figure~2. The best fit slope
of the underlying hard power law is $\Gamma = 2.1$, while $\Gamma = 3.7$ for the 
soft excess power law. The flux at 1 keV is $5.4 \times 10^{-4}$
photons/cm$^{2}$/s/\AA\ in the $\Gamma = 2.1$ power law and $3.3 \times 10^{-6}$ photons/cm$^{2}$/s/\AA 
in the $\Gamma = 3.7$ power law. 

Overall, the broad spectral features in the DWA model
are too shallow compared to the data. The depth of the edges could be increased
if the flux in the underlying power law were higher, but the overall slope
of the resulting spectrum would not match the data.  
A combination of the \ion{Fe}{1} edge
and the n=2-3 lines near 16 \AA\ (Behar et al. 2001) causes the 
flux to rise between 17\AA\ and 19\AA, but not as much as observed in the data.
The rise in X-ray flux between $\sim$22\AA\ and $\sim$24\AA\ is not 
well
reproduced by a combination of the Galactic neutral oxygen K edge and \ion{O}{1}
absorption in the Seyfert, and neither does the model reproduce the rise
in flux between  $\sim$31\AA\ and $\sim$33\AA. 

The best
fit neutral column densities are summarised in Table~2. The species with the
largest neutral column densities are carbon and iron. However, the neutral 
carbon edge is
not within the RGS wavelength range. 

Given the size of the neutral iron edge that is fitted, we would expect to see a much stronger
\ion{Fe}{1} absorption line at $\sim$17.7\AA\ in the observed frame 
(see Figure~2). 
To measure the observed strength of this line as accurately as possible, we have 
modelled the continuum in 
the 16\AA\ to 22\AA\ region with a fifth order polynomial, together with absorption lines
due to \ion{O}{7}, \ion{O}{8} and \ion{Fe}{1}. Use of a local polynomial continuum
minimises systematic uncertainties due to inadequacies in the global model.
The \ion{Fe}{1} line is blended with
\ion{O}{7} 1s-5p, and we measure a combined equivalent width of $38\pm 9$m\AA. Ascribing
half of this, roughly $20$m\AA, to \ion{Fe}{1}, we derive an \ion{Fe}{1} column
of $2.3 \pm 0.4 \times 10^{16}$ cm$^{-2}$ for a Doppler b width of 100 km/s. The
line is on the linear part of the curve of growth, and it is also 
an unresolved transition array (UTA; Behar et al. 2001)
with an effective width of 600 km s$^{-1}$. Thus the derived column density is insensitive
to uncertainties in b, and only increases to $2.9 \pm 0.5 \times 10^{16}$ cm$^{-2}$ 
if we assume b=50 km/s. This is a factor of 5-7 times less than the column density
required by the fits to the neutral iron edge in the DWA model (Table~2). 
Conversely, based on the measured strength of the 17.7A
line, the neutral iron edge is expected to be weak and can
not explain the large 17A continuum break.

\subsubsection{Relativistic Emission Lines}

The second model seeks to fit the soft X-ray spectral features with emission lines
from an ionized, relativistic accretion disk (BR01, Laor 1991). Relativistic broadening
has been invoked previously to explain the profile of the iron emission line near 
6 keV observed first from MCG-6-30-15 using ASCA (Tanaka et al. 1995; see also 
Lubi\'nski \& Zdziarski 2001 and references therein).  Following BR01 
we fit the RGS spectrum with emission lines due to H-like Ly$\alpha$ lines of \ion{O}{8}, \ion{N}{7}
and \ion{C}{6}. 
The model includes a warm absorber component with identical atomic data to that
used in the DWA fit. As in the DWA model, we allow the column density of each warm
absorber ion to vary to get the best fit, and assume b=100 km/s. 
Unlike the DWA model, only a single power law continuum is 
required in the REL fit.
\renewcommand{\arraystretch}{.6}

\begin{deluxetable}{lllll}
\tablewidth{0pc}
\tablecaption{Relativistic Emission Lines}
\tablehead{
\colhead{Parameter}  & \colhead{\ion{C}{6}} & 
   \colhead{\ion{N}{7}}  & \colhead{\ion{O}{8}} & \colhead{Fe K} \\
}
\startdata
E(keV: C,N,O fixed)& \hfil 0.3675  & \hfil 0.5003 & \hfil 0.6536 & \hfil 6.73$\pm$0.07\\
R$_{in}$ (R$_G$) & 1.24$^{+0.45}_{-inf}$ & 1.24$^{+0.30}_{-inf}$ 
    & 1.24$^{+0.25}_{-inf}$ & 1.32$^{+0.80}_{-0.09}$ \\
Emissivity Index & 3.79$^{+0.28}_{-0.06}$ & 3.73$^{+0.07}_{-0.04}$ 
    & 3.51$^{+0.04}_{-0.02}$ & 3.53$^{+0.19}_{-0.17}$ \\
Flux (10$^{-3}$ ph s$^{-1}$ cm$^{-2}$) & 2.99$^{+0.12}_{-0.67}$ 
  & 3.65$^{+0.06}_{-0.23}$   & 3.71$^{+0.03}_{-0.19}$ & 0.12$\pm$ 0.1 \\
EW (eV) & 42$^{+2}_{-9}$ & 96$^{+2}_{-6}$ 
    & 169$^{+1}_{-9}$ & 684$\pm$55 
\enddata
\tablecomments{fixed parameters: redshift=0.012929, disk inclination 35$^\circ$, 
disk outer radius =400 R$_G$. Errors are 90\% confidence.}
\end{deluxetable}

\begin{deluxetable}{lll}
\tablewidth{0pc}
\tablecaption{Warm absorber: Equivalent hydrogen column densities}
\tablehead{
\colhead{Element (Ion)}  & \colhead{N$_{\rm H}$(DWA)} &  \colhead{N$_{\rm H}$(Rel)} \\
}
\startdata
C(V-VI) & \hfil 1.9 & \hfil 1.6 \\
N(VI-VII) & \hfil 1.2 & \hfil 1.6 \\
O(VII-VIII) & \hfil 0.33 & \hfil 0.35 \\
Ne(IX-X) & \hfil 1.3 & \hfil 0.66 \\
Fe(VII-XXIV) & \hfil 18.9 & \hfil 10.7
\enddata
\tablecomments{units are $10^{20}$ cm$^{-2}$.}

\end{deluxetable}

The contribution of the individual components that make up the 
Relativistic Emission Line (REL)  
model is illustrated
in the lower panel of Figure~2. Overall,
the best fit REL model (Figure~4) has a 
$\chi^2$/d.o.f.=1.7, significantly better than the DWA fit. The REL model is
more successful in reproducing the overall shape of the observed
spectrum, and consequently the 
systematic deviations of the residuals plotted in Figure~4
are less marked than 
those of the DWA fit. The characteristic sawtooth rises in flux at 
$\sim$18\AA, $\sim$23\AA\ and $\sim$32\AA\ are modelled by the overlapping
asymmetric profiles of \ion{O}{8}, \ion{N}{7} and \ion{C}{6} emission lines. The
line energies are consistent with the Ly$\alpha$ transitions of these ions
as reported by BR01. 

While visibly a better fit, the formal value of 
$\chi^2$ for the REL model is still high. There is a significant contribution 
to the reduced 
$\chi^2$ of both the
REL and DWA fits from inner shell resonances of low ionization species 
(NI-V, OI-VI, etc), which we have not yet included in either model. These do not affect 
the relative comparison between them however.

We have assessed the contribution of other potential disk emission lines to
the observed spectrum, particularly \ion{O}{7} which is predicted in the
accretion disk models of Ballantyne, Ross \& Fabian (2002). Assuming the same
profile as the \ion{O}{8} line, we can place a 3$\sigma$ upper limit on the
flux of \ion{O}{7} relative to \ion{O}{8} of $\sim$0.11. 


The best fit continuum slope for the REL model has $\Gamma = 2.15$, with 
a flux at 1 keV of $6.8\times 10^{-4}$
photons/cm$^{2}$/s/\AA.
The line profiles are consistent with 
a common set of accretion disk parameters as listed in Table~3, and 
are indicative of emission from just outside the last stable
orbit about the black hole (R$_i=$1.24 R$_{\rm G}$). The emissivity index
of the accretion disk is steep ($\sim$3.7) and is similar for the three ions. This
means that 86\% of the disk emission occurs within 6 R$_{\rm G}$ of the black hole.
The lines are too broad to be explained by emission outside the last stable
orbit of a non-rotating (Schwarzschild) black hole.  The blue shoulders of the model 
lines have a peak-to-base width of
about $\Delta\lambda/\lambda=0.05$ 
which is equivalent to $\sim$33 eV (0.9\AA) for \ion{O}{8}. This 
corresponds to an accretion disk inclination of 35$^\circ$.



\subsection{Spectral Variability}

To search for variability 
of the RGS spectral features with source intensity,
we have divided the data according to whether they are
taken around a peak of the 1-2 hour timescale flares in the light
curve, or in the interflare period. The time intervals used are marked 
in Figure~1.
We have not used data from the low state that occurs towards the end of the 
observation in the variability analysis. This is because
the total number of source counts is significantly less in this time period.
Moreover there is an increase in the background
associated with the end of the spacecraft orbit
that further degrades the statistical
precision of the data during the low-flux interval.
The individual `flare' and `interflare' spectra are plotted in Figure~5, where
we have binned the data in wavelength to increase signal to noise. The ratio of the
two spectra are shown in the upper plot of the Figure.

Overall, the high- and low-flux spectra have a similar shape and show the
same spectral features. The 
overall ratio is relatively 
flat, supporting our earlier assertions that the flux variability in Mrk~766 
is not associated
with a gross change in spectral shape. The $\chi^2$ for a fit of the ratio to a
horizontal line is 151 for 123 degrees of freedom. Nevertherless there are 
significant deviations 
from 
the mean at about the $\pm$15\% level. The most prominent are a peak in the ratio at about 12\AA, and
a dip at about 22\AA\ accompanied by an excess with
similar significance to longer wavelengths. In each feature the maximum deviation of individual points
from the mean is
only slightly more than 3 sigma, but the deviations are systematic over a number of 
neighbouring
points. The excess ratio near 12\AA\ may be due to a complex of \ion{Fe}{22} and 
\ion{Ne}{10} emission. The centroid of the dip at 22\AA\ (22.2\AA) is close to 
the wavelength of
low-level \ion{O}{7} forbidden line emission seen in the spectrum (22.38\AA\ in the
observed frame; see section
3.3) but the width and flux of the forbidden line are insufficient to account for
the feature we see, and its centroid is significantly different. 
The 22\AA\ feature coincides with the blue wing of the putative broad \ion{N}{7} line,
and it may be that we are seeing changes in either the blue wing of the \ion{N}{7} line
or the red wing of the \ion{O}{8} line during flares. Interestingly, there may also
be a lower significance feature shortward of the \ion{C}{6} Ly$\alpha$ rest wavelength
at about 31\AA, but the reality of this feature is less certain. 
There is no significant change in ratio associated with the blue wing of the \ion{O}{8} line.

\subsection{Warm Absorber}

The presence of a warm absorber is betrayed by the numerous narrow absorption lines 
of ionized species in the RGS spectrum of Mrk~766. As noted previously, our
approach is to model these lines empirically as we seek to understand the nature
of the underlying continuum. In Table~1 we list the column density of individual 
ions for both the DWA and REL models. These column densities are naturally only
meaningful if the underlying model accurately reproduces the continuum in the 
vicinity of the relevant line, so generally the values for the REL model are
most representative since this is a better approximation to the observed continuum. 
We also stress that the column densities will be 
sensitive to the Doppler line width assumed (see section 3.1.1). Furthermore the
absorption lines are likely to be filled in to some extent by emission
from the extended warm absorber gas. 

There is indeed weak \ion{O}{7} forbidden emission in Mrk~766 near 22\AA\ (Figure~8) and 
this can 
be used to roughly estimate the covering fraction of the warm absorber by comparing it 
to the measured flux absorbed by the \ion{O}{7} 1s-2p resonance line. We measure a flux
in the forbidden emission line (f in Figure~8) of $1.5 \pm 0.8 \times 10^{-5}$ ph cm$^{-2}$ s$^{-1}$,
and absorbed flux in the
\ion{O}{7} 1s-2p resonance (r) line of $1.9 \pm 0.7 \times 10^{-5}$ ph cm$^{-2}$ s$^{-1}$. 
The ratio of the forbidden line emission flux to 
the corresponding
{\it emission} in the
2p-1s line 
can be estimated based on measurements of NGC~1068, where Kinkhabwala et al. (2002)
find a ratio of 2.1. Thus, assuming similar densities for the warm absorbers in
Mrk~766 and NGC~1068, we expect emission from \ion{O}{7} 2p-1s at a level
of $0.71  \times 10^{-5}$ ph cm$^{-2}$ s$^{-1}$. This means that 30\% of the 
\ion{O}{7} 1s-2p absorption line is expected to be filled by emission and 
the
corrected absorption line flux is therefore 
$\sim 2.6 \pm 0.7 \times 10^{-5}$ ph cm$^{-2}$ s$^{-1}$.
The ratio of resonantly scattered to total flux in the 1s-2p emission line in NGC~1068
is found to be 0.44. 
If we assume that this is also representative of Mrk~766, 
we derive an estimate of about 12\% for the covering factor of the warm absorber 
in the latter. This number should be treated with caution, since in practice 
the scaling from NGC~1068 is likely to be sensitive to differences in
the mean column density and clumpiness of the warm absorber in the two sources.
Nevertheless this exercise demonstrates that the 
implied warm absorber covering factor in Mrk~766 is at least reasonable.

In Table 4 we list the equivalent hydrogen column densities of the 
principal line-forming elements in the warm absorber, 
summed over the ionic species measured, and assuming cosmic abundance. As before these
values assume b=100 km/s.
The implied 
column of oxygen in the warm absorber is a factor of 
four lower than either carbon or nitrogen while the column density derived from the 
iron lines is an order of magnitude 
higher than from the other measured elements. 
We note in passing that the relative strength of the broad \ion{N}{7} and \ion{O}{8} emission 
lines fitted in the REL model also
requires that the nitrogen
abundance relative to oxygen is higher than the solar value by a similar factor, 
as noted by BR01. 
The iron column in the warm absorber is summed over a much 
larger range of 
ionisation states than the other elements and it is possible that carbon, nitrogen 
and oxygen exist largely in ionisation states that we 
have not measured. Strictly speaking, a physical ionisation model of the warm absorber gas is 
required to interpret these column densities
convincingly in terms of elemental abundances. However, the observed overabundance 
of iron, by an order of magnitude, is 
rather large to be explained away in this manner. We note that a similar overabundance
of iron has been measured in NGC~3783 (Blustin et al. 2002) and IRAS~13349+2438
(Sako et al. 2001). 

We should recognise however that these values are likely to be sensitive to uncertainties
in the Doppler width, b, of the lines, and that the sensitivity will be different for
different ions. For example, when we use b=65 km/s the column density of carbon is
about three times higher, with lesser increases for nitrogen
and oxygen. The column density of iron is less sensitive to b, however, since many
of the iron features are UTAs (see section 3.1.1). Thus a decrease in b would serve to
decrease the abundance discrepancy between iron and the lighter elements.

\subsection{MOS and RGS data}

The EPIC MOS-2 data are shown in Figure~6.  The sawtooth shaped spectral features 
that are shown to advantage by the RGS data are
also clearly
present in the MOS.
We have fitted the data with the REL model 
described above,
simultaneously with the RGS data, adding a fourth relativistically broadened
line at 6.7 keV to represent iron emission.  The best fit power law 
continuum parameters are indistinguishable
from those that were derived from the RGS data
alone. 
While this model is a reasonable representation of the data over most of the energy range,
a simple power-law does underpredict the flux above the iron line by an
amount that rises to 30\% at 10 keV. Because
the response of the instrument falls off towards high energies, this excess has
relatively little statistical weight in the overall fit however. We can model the excess
with a power-law that has a break above 5.6 keV, hardening by $\Delta\Gamma=0.4$ 
to accommodate the high energy excess.
This improves the reduced $\chi^2$ from 2.2 to 2.1 per d.o.f for 1624 d.o.f. 
The broken power-law model fit is illustrated in 
Figure~6. It is likely that the 
high energy excess is due to reflection from a disk. Indeed we can represent the spectrum
equally well (i.e. identical $\chi^2$) with a model (developed by Piotr Zycki) that includes 
reflection from a constant density X-ray illuminated atmosphere.  It
computes the Compton-reflected continuum (cf. Magdziarz \& Zdziarski, 1995) and 
the Fe K$\alpha$ line 
(cf. Zycki \& Czerny, 1994) and adds relativistic smearing around a Schwarzschild black hole.

The reduced $\chi^2$ of the best-fit REL model is still formally poor, at 2.1 per d.o.f.
However, as evidenced by Figure~6, the systematic deviations of the data from
the model are small, and the poor reduced $\chi^2$ is largely due to unmodelled
low-level spectral complexity, particularly at the lowest energies.
As with the RGS data on their own, the DWA model (together with a relativistic 
Fe K$\alpha$ line and high energy excess) is a significantly poorer fit 
to the combined
MOS and RGS data than the REL model, yielding $\chi^2$/d.o.f=3.0. 
Significantly, even over the extended energy range
covered by the MOS,
the REL model still requires only a single 
underlying $\Gamma = 2.15$ power law continuum to fit the data, 
provided we include a disk reflection component, or its equivalent, to account for the
high energy excess. 

In Figure~7 we plot the residuals of the MOS-2 data in the 3-8 keV range 
with respect to the best
fitting broken power-law continuum. We use this simple continuum rather than the
full reflection spectrum model 
to illustrate the actual shape of the observed data in the vicinity of the iron line 
without introducing
model-dependent line-like features. We also show the REL model profile (Laor 1991) 
used to fit the iron line in the broken power-law model, the parameters of which are listed in Table~3.
The data support the notion of a strongly asymmetric line
whose red wing extends down to about 3 keV. Its energy is
indicative of highly ionized iron, and the profile 
is consistent with the
same set of disk parameters used for the low-energy lines (Table~3).
The detection of the red wing in 
the iron line 
confirms the report 
by Page et al. (2001). 
In the lower panel of Figure~7
we have scaled the MOS-2 iron line profile to the energy of 
the \ion{O}{8} line and superimpose this on the RGS data. There is good agreement
between the MOS and RGS profiles, quantified by the conjunction of the model
line parameters and lending credence to the idea that the RGS spectral features
are caused by relativistic emission lines.  

\section{Discussion}

The long observation of Mrk~766 reported in this paper, which sampled the source
at a higher mean flux level than previously, provides the opportunity to
test the rival models that have been put forward to explain the low-energy
spectral features first reported by BR01. We have sufficient 
signal-to-noise to
constrain the variability of these features as a function of source flux. 
We have also used the EPIC instrument on XMM-Newton to confirm the presence
of a broad, asymmetric iron line feature at 6.7 keV, and use the
combination of the RGS and EPIC instruments on XMM-Newton to compare the
iron line profile with the low-energy spectrum. 

A point of debate in 
the literature has been the 
discontinuity in the spectrum of Mrk~766 at
about 18\AA. A similar feature has also been seen in MCG-6-30-15 
by BR01 and Lee et al. (2001). 
Conceivably it is an absorption edge. 
Its wavelength, however, is inconsistent with the O{\sc VII} edge expected 
from a warm absorber
unless the material producing it is redshifted by an implausible amount 
($\sim$ 16,000 km/s). Thus Lee et al. (2001) propose 
that
the feature in MCG-6-30-15 is due to the L-edge of neutral iron, which 
matches the 
observed wavelength more closely than the O{\sc VII} edge. 
When interpreted in this way, the depth of the
feature in MCG-6-30-15 implies an equivalent hydrogen column of 
$N_H = 4 \times 10^{21}$ cm$^{-2}$ 
assuming solar abundances. This cannot be a column of neutral gas, 
otherwise the soft
X-ray spectrum would be highly absorbed by C{\sc I} and O{\sc I} edges, which
is not seen. Besides, the presence of numerous narrow absorption
lines from ionized species is indicative of a warm absorber. This leads Lee et al.
to conclude that the neutral fraction of the absorber, particularly iron, 
is bound up in dust
that is immersed in the ionized medium. Such dusty warm absorber (DWA) models have 
previously been invoked to explain the apparent discrepancy
between the large reddening of certain AGN deduced from their optical-UV spectrum,
and the lack of corresponding absorption of the X-ray spectrum (e.g. Reynolds et al. 
1997; Walter \& Fink 1993). 

We have developed a DWA model and have fitted it to our RGS data on Mrk~766.
The model struggles to reproduce the overall shape of the RGS 
spectrum. If we nevertheless interpret the sharp discontinuity 
in flux near 18\AA\ 
as being due to neutral iron, 
the column 
density implied is $1.7 \times 10^{17}$ cm$^{-2}$. 
This is of the right order to be consistent with the estimated
reddening of Mrk~766 (E(B-V)=0.4; Walter \& Fink 1993) for reasonable abundances. 
Given the required depth
of the Fe I L edge we would, though, expect to see much stronger 
Fe I L resonance lines
than are observed. 
Furthermore, there is no compelling evidence for K edges of 
other neutral atoms such
as \ion{O}{1}, \ion{Mg}{1} \& \ion{Si}{1} at the redshift of Mrk~766, and the
composition of the putative dust grains is therefore unphysical. 
The DWA fit also requires a significant column of neutral carbon, forced by the slope of
the long-wavelength spectrum in the RGS data. The carbon 
edge itself is not within the RGS range, but the MOS data extend to these energies and 
beyond. Even accounting for uncertainties in the MOS response matrix at these energies,
we find no evidence for the expected increase in flux below 0.2 keV, which should
amount to a factor of two if the edge is present at the level implied by the
fits to the RGS data.  

In contrast, the relativistic emission line (REL) model is more successful 
in reproducing the 
Mrk~766 RGS 0.3-2 keV data. The broad spectral features are consistent with 
Ly$\alpha$ emission lines of \ion{O}{8}, \ion{N}{7} and \ion{C}{6} at the 
rest wavelength of the galaxy, superimposed on an underlying power-law. An 
extension of this model including a fourth line to represent emission due to iron 
at 6.7~keV and reflection from a disk, also fits the 0.2-10 keV EPIC MOS data 
on Mrk~766. The profiles
of the low energy emission lines in Mrk~766 are consistent with that measured
for the iron emission at 6.7 keV.
The REL model requires only 
a single underlying power law continuum (with the excess above 7 keV being
accounted for by reflection), unlike the DWA model where  
a steep low-energy excess 
component below about 0.7 keV is also needed.

Thus, adding dust to a warm absorber is not a convincing description of the Mrk~766 
X-ray
spectrum while
the REL model, suggested for Mrk~766 originally by BR01 to explain the 
earlier, shorter XMM-Newton
observation, has become more compelling when
confronted with the higher signal to noise data reported here. 
In many ways Mrk~766 is a more clear-cut example of the phenomenon than MCG-6-30-15 
because the spectrum is
less affected by the warm absorber.

The underlying physics that might produce such a (REL) spectrum is still controversial. 
Ballantyne \& Fabian (2001) fit a model to the ASCA spectrum of MCG-6-30-15 
that involves reflection of 
hard X-rays from ionized layers on the surface of an accretion disk. They optimise
the model parameters to 
best fit the iron K line, but this 
falls short of
reproducing the observed equivalent width of the low-energy lines by more 
than an order of
magnitude. For example, their model predicts an equivalent width of only 6eV for the
oxygen line assuming solar 
abundance. In the Ballantyne \& Fabian  model, however,  
the iron K line is a mix of emission from 
an ionized inner disk, and a sharper neutral component from further out. 
It is not clear how well the ionisation state of the inner disk region is
being constrained in the model. 

In more recent work, Ballantyne, Ross \& Fabian (2002) 
relax the requirement to fit the iron line and thus widen the
scope of parameter space explored. In this case there are regimes in which 
their model does predict
equivalent widths for the low energy lines comparable to those that we measure.
Sako et al. (2002) also describe a thermal/ionization structure
for the accretion disk surface layers in which Ly-$\alpha$ recombination lines
of appropriate equivalent width could plausibly be produced. 
The lines will be intrinsically broadened by Compton scattering in the
hot disk atmosphere. However, the widths that Ballantyne et al. predict for the blue wing 
of the lines 
(a few tens of eV) 
are comparable to those that we measure for the underlying 
emission lines in our Mrk~766 data, when the
effects of absorption line and edge features in the warm absorber are taken into account.
The Ballantyne et al. models predict additional weaker line features due to 
\ion{O}{7} and \ion{N}{6} that are not required by our fits. However, the upper limit 
that
we can set on an \ion{O}{7} line in our data, measured with respect to the 
observed flux
of \ion{O}{8}, is only marginally inconsistent with
the predictions of the Ballantyne et al. model for an ionizing parameter ($L/nr^2$) 
of 250 erg cm s$^{-1}$
and Solar oxygen abundance. 
Sako et al. (2002) in any case
use a different ionisation structure for the hot layer, in which these lines are 
supressed. Sako et al. also argue that iron L-lines above the \ion{O}{8} K edge at 
14.22\AA, predicted by Ballantyne et al., will be absorbed and the 
energy re-radiated at \ion{O}{8} Ly-$\alpha$.

We have searched for evidence of variability in the low-energy spectral
features of Mrk~766 in response to the 1-2 hour timescale flares
that are seen in the light curve. There are indications of a systematic change in the high-to-low
flux ratio at the $\pm$15\% level particularly in the region between about 22\AA\ and 24\AA. 
This signature is tantalising, since, in the context of the REL model, it may
indicate that either the blue wing of the N line or the red wing of the O line are responding 
to the flares, which
would be an important physical constraint on models of the emission region. However the
current statistical significance of the effect is low. Confirmation
of these changes with higher signal to noise data will be important. 


Clearly more work is required before we fully understand 
the low energy X-ray spectrum of Mrk~766 and similar
objects. On the observational side, there are a number of directions for
future research. The variations in the profile of the lines, if that is what
they are, with source flux are at the limits of detection in the current RGS data.
Confirmation in other observations would provide an important physical constraint
on potential models. We intend to examine the issue of spectral variability
in Mrk~766 in more detail in a future publication, including consideration of the
EPIC data. Similarly, a 
physical model of the warm absorber gas could be valuable in distinguishing
which features come from the warm gas and which have other causes.
It is encouraging, however, that the detail revealed by the current RGS data, 
coupled with the high throughput 
and energy coverage of the EPIC cameras, gives us real discriminating power. 
We have shown that the
REL model is an elegant and relatively simple description of the 0.3-10 keV spectrum
of Mrk~766. 
The physical processes that 
might produce such lines clearly demand further 
consideration as a mechanism to account for the spectrum of this and other
Seyfert galaxies.

\acknowledgments
We thank Ted Brookings (UCSB) and Jelle Kaastra (SRON) for their work in developing 
the IMP and SPEX  spectral 
fitting codes used in this analysis.

\end{document}